\documentclass[twocolumn,showpacs,preprintnumbers,aps,pra,amsmath,amssymb,floatfix,superscriptaddress]{revtex4-1}

\usepackage{graphicx}
\usepackage{dcolumn}
\usepackage{bm}
\usepackage{color}
\usepackage{lineno}
\usepackage{natbib}
\usepackage{hyperref}
\hypersetup{colorlinks,breaklinks,urlcolor=blue,linkcolor=blue,citecolor=blue,filecolor=blue}

\begin{document}

\title{Observation of resonant scattering between ultracold heteronuclear Feshbach molecules}

\author{Fudong Wang}
\email{wangfd@sustech.edu.cn}
\affiliation{Department of Physics, The Chinese University of Hong Kong, Hong Kong, China}
\affiliation{Shenzhen Institute for Quantum Science and Engineering and Department of Physics, Southern University of Science and Technology, Shenzhen, 518055, People's Republic of China}
\author{Xin Ye}
\affiliation{Department of Physics, The Chinese University of Hong Kong, Hong Kong, China}
\author{Mingyang Guo}
\thanks{Current address: 5{.} Physikalisches Institut and Center for Integrated Quantum Science and Technology, Universit{\"a}t Stuttgart, Pfaffenwaldring 57, 70569 Stuttgart, Germany}
\affiliation{Department of Physics, The Chinese University of Hong Kong, Hong Kong, China}
\author{D. Blume}
\affiliation{Homer L. Dodge Department of Physics and Astronomy, The University of Oklahoma, 440 West Brooks Street, Norman, Oklahoma 73019, USA}
\affiliation{Center for Quantum Research and Technology, The University of Oklahoma, 440 West Brooks Street, Norman, Oklahoma 73019, USA}
\author{Dajun Wang}
\email{djwang@cuhk.edu.hk}
\affiliation{Department of Physics, The Chinese University of Hong Kong, Hong Kong, China}
\affiliation{The Chinese University of Hong Kong Shenzhen Research Institute, Shenzhen, China}

\date{\today}

\begin{abstract}
We report the observation of a dimer-dimer inelastic collision resonance for ultracold Feshbach molecules made of bosonic sodium and rubidium atoms. This resonance, which we attribute to the crossing of the dimer-dimer threshold with a heteronuclear tetramer state, manifests itself as a pronounced inelastic loss peak of dimers when the interspecies
scattering length between the constituent atoms is tuned. Near this resonance, a strong modification of the temperature dependence of the dimer-dimer scattering is observed. 
Our result provides insight into the heteronuclear four-body system consisting of heavy and light bosons and offers the possibility of investigating ultracold molecules with tunable interactions. 

\end{abstract}

\maketitle

\section{Introduction}
\label{sec_Intro}

Feshbach resonances (FRs), which occur when the energy threshold of two colliding atoms is tuned to coincide with the energy of a bound state in another scattering channel, have 
been widely used in ultracold atom systems~\cite{Kohler06,Chin2010}. In the vicinity of a FR, the interaction between two atoms can be changed almost at will by tuning the strength of an external magnetic field. In addition, FRs can also be used to convert atom pairs into weakly-bound Feshbach molecules (FMs). For instance, for two-component quantum degenerate Fermi gases, long-lived FMs can be created by sweeping the magnetic field across a FR. This technique has been the workhorse for the study of the BEC-BCS crossover physics, which has deep connections with high-$T_c$ superconductivity and a myriad of other exotic many-body phenomena.

Feshbach molecules can also be created out of ultracold bosons. The resulting dimers are prone to inelastic losses due to the absence of Pauli blocking~\cite{Mukaiyama2004,Thompson2005,Syassen2006} and thus much less studied. However, it was pointed out  for single-species bosons~\cite{DIncao2009,Deltuva2011} that these dimers have intriguing scattering properties near unitarity that are connected with elusive weakly-bound tetramer states. By varying the strength of an external magnetic field, the dimer-dimer scattering threshold can be tuned to be energetically degenerate with one of the tetramer states, causing  Feshbach-like resonances. Around these resonances, the dimer-dimer scattering length can be tuned from positive to negative values, although the atom-atom scattering length is always positive~\cite{DIncao2009,Deltuva2011}.     

Weakly-bound tetramer states are of great interest in the context of few-body physics in connection with the celebrated three-body Efimov effect~\cite{Efimov1970,Kraemer2006,Braaten2006, Naidon2017} and its extension to systems with more particles~\cite{Hammer2007,Stecher2009}. With identical bosons, signatures of four-body resonances were observed through atom losses~\cite{Ferlaino2009,Pollack2009}. In the homonuclear Cs$_2$ gas, Feshbach-like loss resonances were observed in a regime where a narrow Cs-Cs $g$-wave resonance sits on top of a broad Cs-Cs $s$-wave resonance~\cite{Chin2005}. Moreover, a loss minimum, rather than a resonance, was reported in the halo-dimer regime~\cite{Ferlaino2008}. For a heteronuclear mixture consisting of light bosons (L) and heavy bosons (H), the situation is more complex since there may exist two distinct heteronuclear trimers (LH$_2$ and L$_2$H) and three distinct tetramers (L$_3$H, L$_2$H$_2$, and LH$_3$)~\cite{Marcelis2008,WangYuJun2012,Blume2014,Schmickler2017,Blume2019}. This heteronuclear scenario is almost entirely unexplored, except for a $^{41}$K$^{87}$Rb dimer experiment where no prominent loss features were detected~\cite{Kato17}. 

\begin{figure}
	\includegraphics[width=0.8\linewidth]{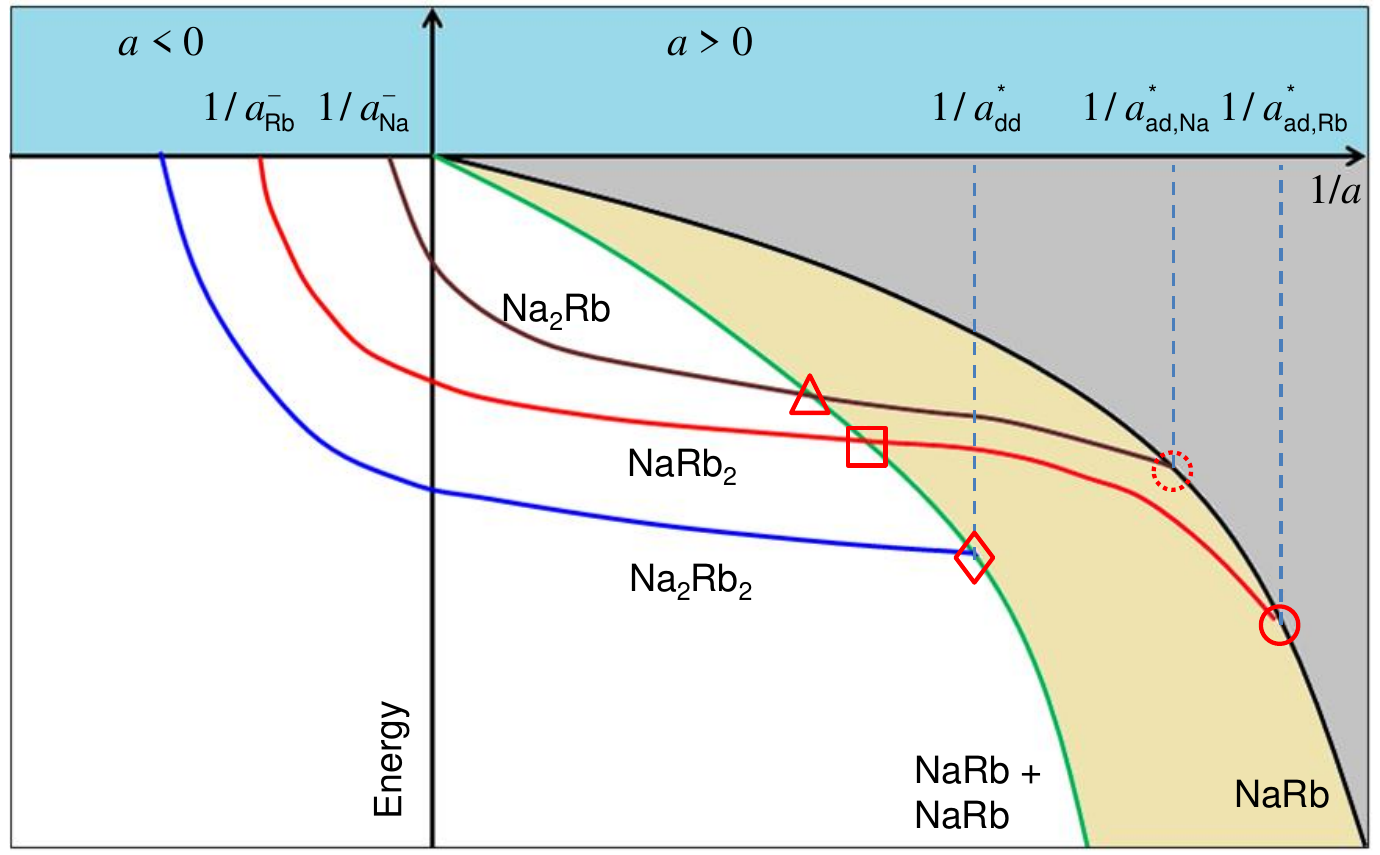}
	\caption{\label{fig1} 
		Schematic few-body energy diagram of the Na-Rb system (Appendix~\hyperref[sec_theory]{\ref{sec_theory}})). The binding energies for one and two NaRb dimers, the NaRb$_2$ and Na$_2$Rb trimers and one of the tetramers, 
		Na$_2$Rb$_2$, are shown. The blue shaded area above zero energy is the atomic scattering continuum, while the yellow and gray shaded areas represent the dimer-dimer and atom-dimer scattering continuum, respectively. The observed resonance for NaRb + NaRb at $a^*_{\rm{dd}}$ is marked by the diamond. The atom + dimer resonances for Rb + NaRb at $a^*_{\rm{ad,Rb}}$ and Na + NaRb at $a^*_{\rm{ad,Na}}$ are marked by the solid and dotted red circles, respectively. The red square and triangle indicate the thresholds of the rearrangement reactions NaRb + NaRb $\rightarrow$ NaRb$_2$ + Na and NaRb + NaRb $\rightarrow$ Na$_2$Rb + Rb, respectively. The corresponding scattering lengths (not marked) are $a_{\rm dd,Na}^{\rm reaction}$ and $a_{\rm dd,Rb}^{\rm reaction}$. On the $a<0$ side, the ``Efimov favored'' NaRb$_2$ and the ``Efimov unfavored'' Na$_2$Rb trimers intersect the three-atom continuum at $a^-_{\rm Rb}$ and $a^-_{\rm Na}$.       
	}
\end{figure}

Figure~\hyperref[fig1]{\ref{fig1}} shows a schematic energy spectrum of the Na-Rb system based on microscopic calculations (Appendix~\hyperref[sec_theory]{\ref{sec_theory}}). A prominent feature of the homonuclear Efimov scenario is the existence of a universal scaling factor $\lambda$ that governs the size and other properties of a series of three-body states~\cite{Efimov1970,Kraemer2006,Braaten2006}. The heteronuclear Efimov scenario~\cite{Braaten2006,WangYuJun2012b}, where the Rb-Rb and Na-Na scattering lengths vanish, is governed by two such scaling factors, one for the ``Efimov favored'' NaRb$_2$ system and one for the ``Efimov unfavored'' Na$_2$Rb system, $\lambda_{\rm{NaRbRb}}=37$ and $\lambda_{\rm{NaNaRb}}=5.1 \times 10^8$, respectively. 
These large scaling factors make it experimentally demanding if not impossible to observe three-atom resonances that are associated with three-body Efimov states on the $a<0$ side. However, on the $a>0$ side, the critical scattering length associated with two NaRb molecules being energetically degenerate with one of the Na$_2$Rb$_2$ four-body states 
(denoted by $a_{\rm dd}^*$ in Fig.~\hyperref[fig1]{\ref{fig1}}) and the critical scattering lengths associated with an atom and a dimer being degenerate with one of the two Efimov trimers 
(denoted by $a_{\rm ad,Rb}^*$ for Rb + NaRb and the NaRb$_2$ trimer, and by $a_{\rm ad,Na}^*$ for Na + NaRb and the Na$_2$Rb trimer, respectively) are more accessible experimentally. 

This work investigates dimer-dimer collisions experimentally with an ultracold sample of $^{23}$Na$^{87}$Rb FMs~\cite{Wangfudong2013,Wangfudong2015b} ($^{23}$Na and $^{87}$Rb will be denoted by Na and Rb hereafter). Tuning the two-body scattering length $a$ between Na and Rb atoms on the $a>0$ side via an interspecies Feshbach resonance, we observed a dimer-dimer loss resonance and attributed it to the crossing of the dimer-dimer threshold and a Na$_{2}$Rb$_{2}$ tetramer state. Near this resonance, we observed a strong modification of the Wigner threshold law with the dimer-dimer collision rate depending non-universally on temperature. 

The organization of this work is as follows: The experimental details for the sample preparation and loss rate fitting are given in Sec.~\hyperref[sec_exp]{\ref{sec_exp}}. In Sec.~\hyperref[sec_dd]{\ref{sec_dd}}, we observe an inelastic resonance in a sample of NaRb Feshbach molecules. The temperature dependent behavior of the loss rate is studied at three different scattering lengths in Sec.~\hyperref[sec_td]{\ref{sec_td}}. Finally, Sec.~\hyperref[sec_con]{\ref{sec_con}} presents the conclusions. Theoretical calculations are relegated to Appendix~\hyperref[sec_theory]{\ref{sec_theory}}.

\section{Experiments}
\label{sec_exp}

\subsection{Sample preparation}
\label{sec_sample}
We start the experiment from an ultracold mixture of Na and Rb atoms both in their lowest hyperfine Zeeman level. The interspecies $s$-wave scattering length is tuned using the Feshbach resonance located at the magnetic field strength $B_0 = 347.64$~G~\cite{Wangfudong2013,Wangfudong2015}. To create FMs, we start at $B= 355$~G and then sweep across the resonance to a final $B$ field of 335.62~G to do the magneto-association. Since the FMs have a vanishingly small magnetic dipole moment near the final $B$ field, the residual Na and Rb atoms can be removed with a strong magnetic field gradient pulse without affecting the FMs~\cite{Wangfudong2013,Wangfudong2015}. Following this procedure, we routinely create pure molecular samples of up to 10$^4$ NaRb FMs. The interspecies scattering length $a$ is then controlled by changing the magnetic field strength between $335.62$~G and $347.48$~G, corresponding to binding energies of the FMs between about $2\pi\times 22$~MHz and $2\pi\times 19$~kHz. These binding energies are comparable to or smaller than the van der Waals energy of $2\pi\times 29.8$~MHz. For detection, the magnetic field is swept reversely across the resonance to dissociate the FMs. The resulting Na atoms are probed using standard absorption imaging methods. The trap frequency that the FMs feel, measured from the sloshing motion, is $\bar{\omega}=2\pi\times76(1)$~Hz; no dependence on the magnetic field strength was detected. The typical sample temperature $T$ is measured to be $343(30)$~nK by adding time-of-flight expansion before dissociating the molecules for detection. The typical initial peak density, calculated from the measured $T$, $\bar{\omega}$ and number of FMs, is $2\times 10^{11}~\rm{cm}^{-3}$.

\subsection{Loss rate measurement}
\label{sec_lossfit}

We measure the loss rate by recording the time evolution of the molecule number after ramping $B$ from $335.62$~G to the desired field strength. In addition to the loss of molecules, an increase of the temperature is observed. The heating is characteristic of inelastic collisions due to the preferential removal of FMs from the highest density part of the sample~\cite{Soding1998,Weber2003}.

In principle, the observed losses can be caused by several possible inelastic processes. Two colliding NaRb dimers can form either a Na$_{2}$Rb or a NaRb$_{2}$ trimer, with the fourth atom carrying away the released binding energy. Alternatively, one of the dimers can relax into a more deeply bound state, with the other dimer being dissociated. Both processes contribute, in principle, to two-body dimer loss~\cite{Marcelis2008}. Unfortunately, our measurements cannot distinguish whether one of these processes is dominant or whether both contribute appreciably. Near the magnetic field strengths at which the loss of dimers is maximal, the magnitude of the dimer-dimer scattering length $a_{\rm{dd}}$ is expected to be large~\cite{DIncao2009,Deltuva2011}. In this regime, the three-dimer recombination may be enhanced in the same manner as three-atom recombination near an atomic Feshbach resonance. Without $a~priori$ assumption, it is not clear which, if any, of these processes dominates the loss of dimers from the molecular sample. However, a fit of the data with maximal loss to a three-body loss model~\cite{Weber2003} leads to a rate coefficient on the order of $2\times 10^{-20}$ cm$^{6}$s$^{-1}$.  This value is orders of magnitude larger than the maximum value of $5.8(1.0)\times 10^{-23}$ cm$^{6}$s$^{-1}$ permitted by the unitary limit~\cite{Chin2005,Rem2013}. It is thus non-physical to designate three-dimer recombination as the main loss mechanism. Although we cannot rule out that two- and three-body losses occur simultaneously, we conclude that the dimer-dimer loss is dominated by two-body processes.

In order to obtain the two-body dimer-dimer loss rate coefficient $\beta_{\rm{dd}}$, we fit the number and temperature evolution simultaneously to two coupled rate equations 
\begin{equation} \label{eq:twobody} 
\begin{split}
\frac{dN}{dt} &=-\beta_{\rm{dd}} \gamma \frac{N^{2}}{T^{3/2}}\\
\frac{dT}{dt} &=\beta_{\rm{dd}} \gamma N \frac{1/4+h_0}{T^{1/2}}.
\end{split}
\end{equation}
Here $\gamma=(\frac{\overline{\omega}^{2} M}{4 \pi k_{B}})^{3/2}$ is a constant with $M=m_{\rm{Rb}}+m_{\rm{Na}}$ the molecular mass and $k_B$ the Boltzmann constant. The $h_0$ term arises from the momentum dependence of $\beta_{\rm{dd}}$, i.e., the variation of $\beta_{\rm{dd}}$ over the thermal distribution~\cite{Soding1998}. 

\begin{figure}
	\includegraphics[width=0.95\linewidth]{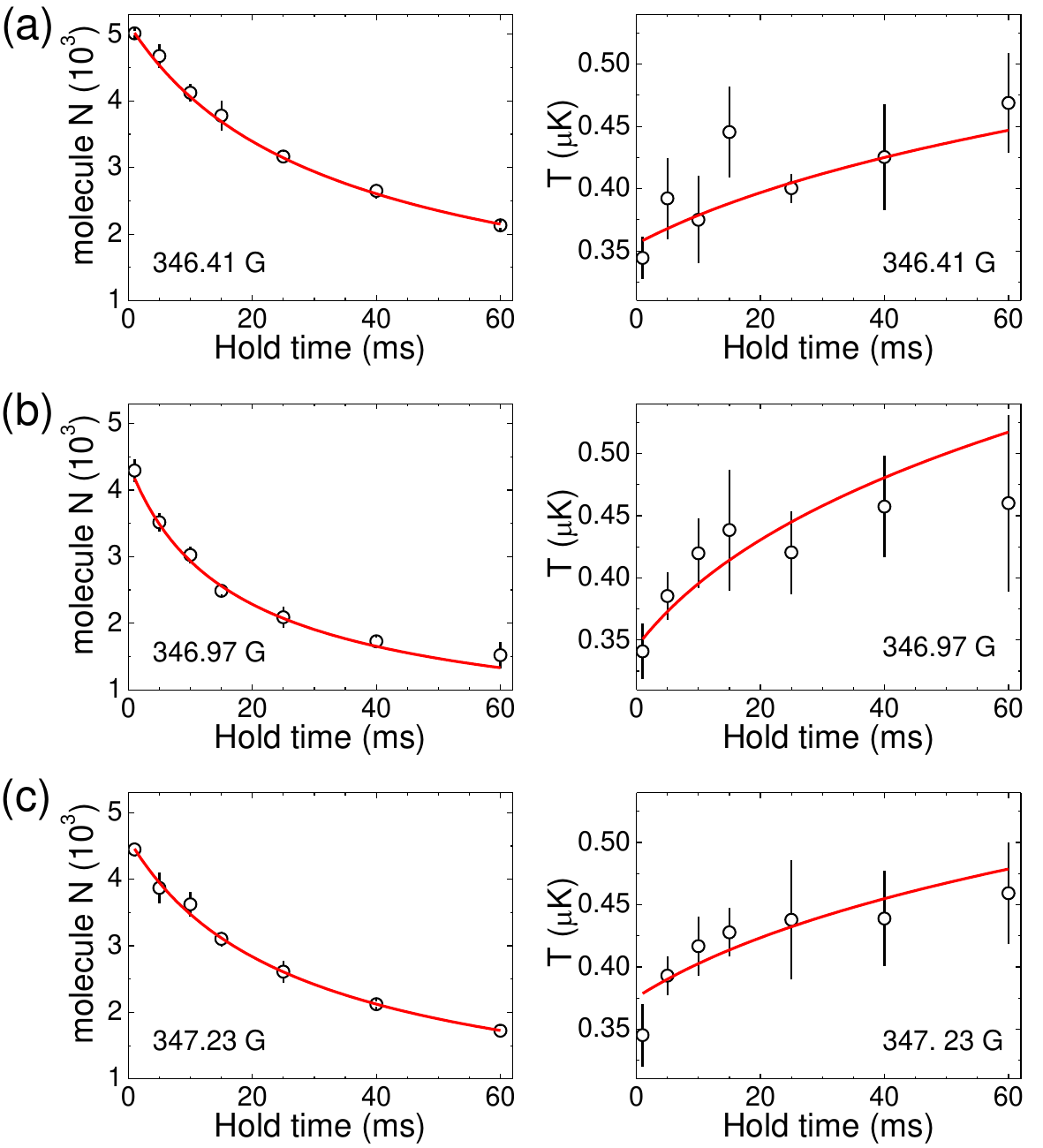}
	\caption{\label{figs2} 
		Inelastic collisions of pure NaRb FMs. Symbols in (a)-(c) show experimentally determined loss (left column) and heating (right column) curves of the FM sample at three different magnetic field strengths, with (a) and (c) taken far from the observed loss peak and (b) very close to it. The red solid curves are fits to the two-body loss model, which is used to extract $\beta_{\rm{dd}}$. }
\end{figure}

Several examples are shown in Fig.~\hyperref[figs2]{\ref{figs2}}(a)-(c). The red solid curves are from fitting to the coupled two-body rate equations (Eq.~\hyperref[eq:twobody]{\ref{eq:twobody}}). Note that samples with nearly the same initial temperatures are used in this measurement. Unfortunately, because of the heating, the temperature $T$ and thus $\beta_{\rm{dd}}$ are actually not constant during the course of the collision. However, the function of $\beta_{\rm{dd}}$ vs. $T$ is unknown. To mitigate this problem, we decided to treat $\beta_{\rm{dd}}$ as a constant in the fitting since during the relatively short hold time of 60 ms, the temperature increases are typically less than 50\% and $\beta_{\rm{dd}}$ should not change dramatically. In measuring the temperature dependence of $\beta_{\rm{dd}}$ (Sec.~\hyperref[sec_td]{\ref{sec_td}}), even shorter hold times of about 25 ms are used to limit the temperature increase to less than 30\%. We note that this treatment inevitably limits the energy resolution of our measurement to the temperature change during the hold time. The energy resolution for the data shown in Fig.~\hyperref[fig_dimerdimer]{\ref{fig_dimerdimer}} is 150 nK, while that of Fig.~\hyperref[fig_varyT]{\ref{fig_varyT}} is shown as the horizontal error bars.


\begin{figure}
	\includegraphics[width=0.9\linewidth]{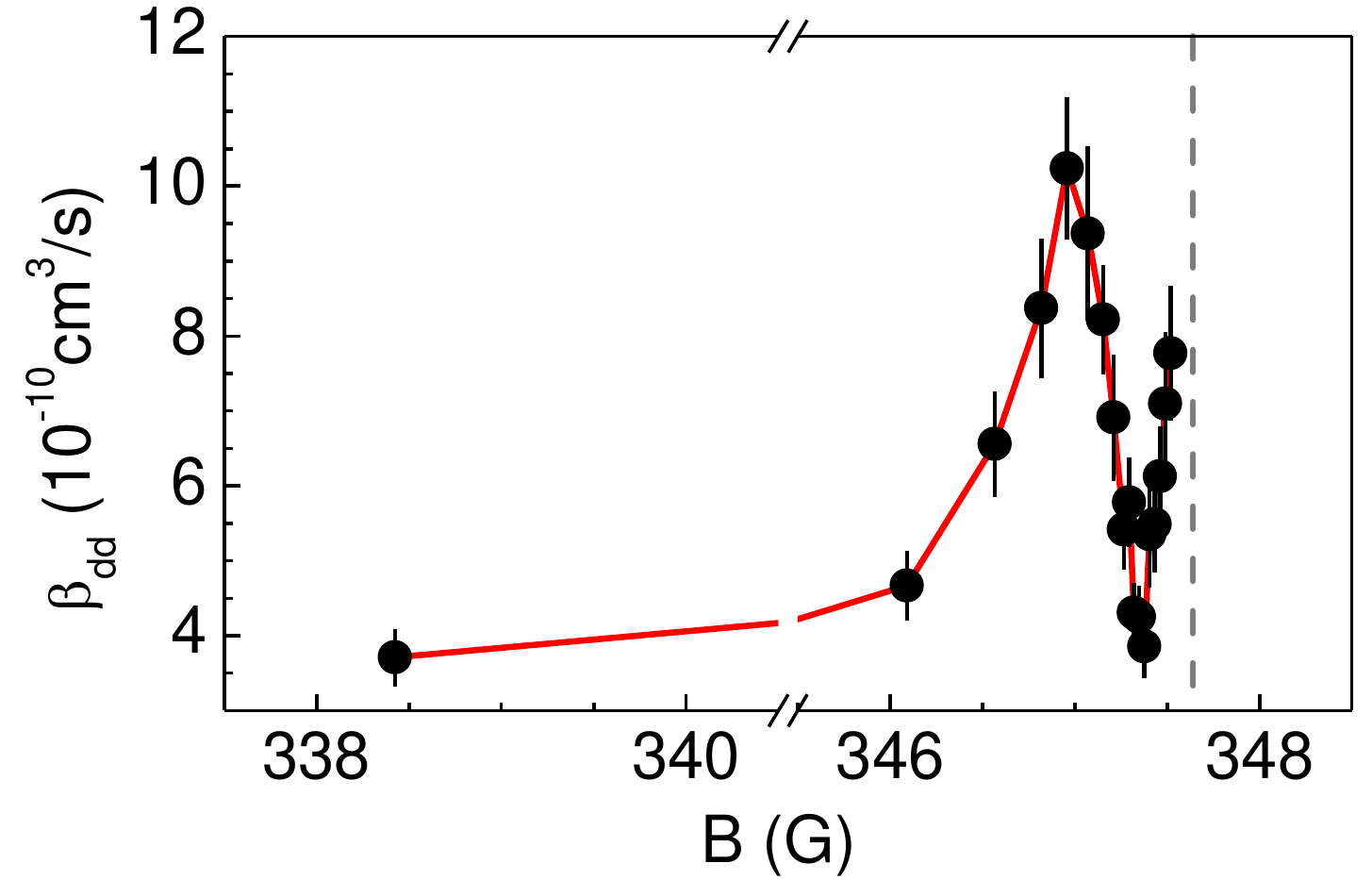}
	\caption{\label{figs3} 
		Dimer-dimer loss rate coefficient $\beta_{\rm{dd}}$ vs. magnetic field strength $B$.
		The position of the interspecies atomic Feshbach resonance is marked by the vertical dashed line. 
		The red curve serves as a guide to the eye. }
\end{figure}

The coupled rate equations assume a Gaussian distribution for a thermalized sample in a harmonic trap. However, even at the loss minimum near $347.38$~G, $\beta_{\rm{dd}}$ is still quite large, namely,  $\beta_{\rm{dd}}=3.7(4)\times 10^{-10}$~cm$^3$s$^{-1}$. Loss rates this large most probably leave the molecules insufficient time to reach equilibrium. This effect, which is neglected in our analysis, could introduce a large systematic error into $\beta_{\rm{dd}}$. Unfortunately, we do not currently have a good way to estimate this possible systematic error.


For the conversion of $B$ and $a$, we use a general expression of two overlapping s-wave resonances \cite{Lange2009}
\begin{equation}
a = a_{\rm bg}\left(1-\frac{\Delta}{B-B_{0}}
\right)\left(1-\frac{\Delta_1}{B-B_{1}}\right),
\label{eq_sc}
\end{equation}
with the background scattering length $a_{\rm bg}=66.8a_0$
, resonant magnetic field strength $B_0 = 347.62$ G and $B_1 = 478.83$ G, and the resonance width $\Delta = 5.20$ G and $\Delta_{1} = 4.81$ G as measured by radio frequency association spectroscopy~\cite{Wangfudong2015}. 

The precise determination of $a$ is essential for the analysis and for connecting experiment and theory. Presently, the $B$-to-$a$-conversion is limited by two factors.
First, the parameters entering into Eq.~(\hyperref[eq_sc]{\ref{eq_sc}}) have uncertainties~\cite{Wangfudong2013}. This can lead to systematic shifts for all values of $a$. Second, because of the relatively small $a_{\rm bg}$ and $\Delta$, the magnetic field needs to be tuned rather close to $B_0$ to obtain large $a$.  
This implies that the short-term magnetic field stability of about $\pm$7 mG translates into a non-negligible uncertainty of $a$ for magnetic field strengths near $B_0$. For the smallest $|B-B_0|$ considered in this work, i.e., for $B=347.480(7)$~G, Eq.~(\hyperref[eq_sc]{\ref{eq_sc}}) yields $a=2320(98)a_0$. 

It is worth mentioning that for the magnetic field strengths considered, the Rb-Rb and Na-Na scattering lengths are approximately constant, $a_{\rm RbRb}=100.4a_0$~\cite{Kempen2002} and $a_{\rm NaNa}=54.5a_0$~\cite{Knoop2011}. As will be discussed in Appendix~\hyperref[sec_theory]{\ref{sec_theory}}, the fact that these scattering lengths are finite and positive has an appreciable effect on the critical scattering length values at which the tetramer and trimers hit the dimer-dimer and atom-dimer thresholds. The universal or semi-universal physics expected at large $a$~\cite{Helfrich2010,WangYuJun2012,WangYuJun2012b,Blume2014} could then be substantially modified. While such modification is well understood in the $^6$Li-$^{133}$Cs Fermi-Bose system~\cite{ulmanis2016heteronuclear,Ulmanis2016},
where only one intraspecies scattering length is non-zero, a quantitative theory for the heteronuclear Bose-Bose case 
that we are dealing with has not yet been developed.


\section{Results: Dimer-Dimer resonance}
\label{sec_dd}

\begin{figure}
\includegraphics[width=0.9\linewidth]{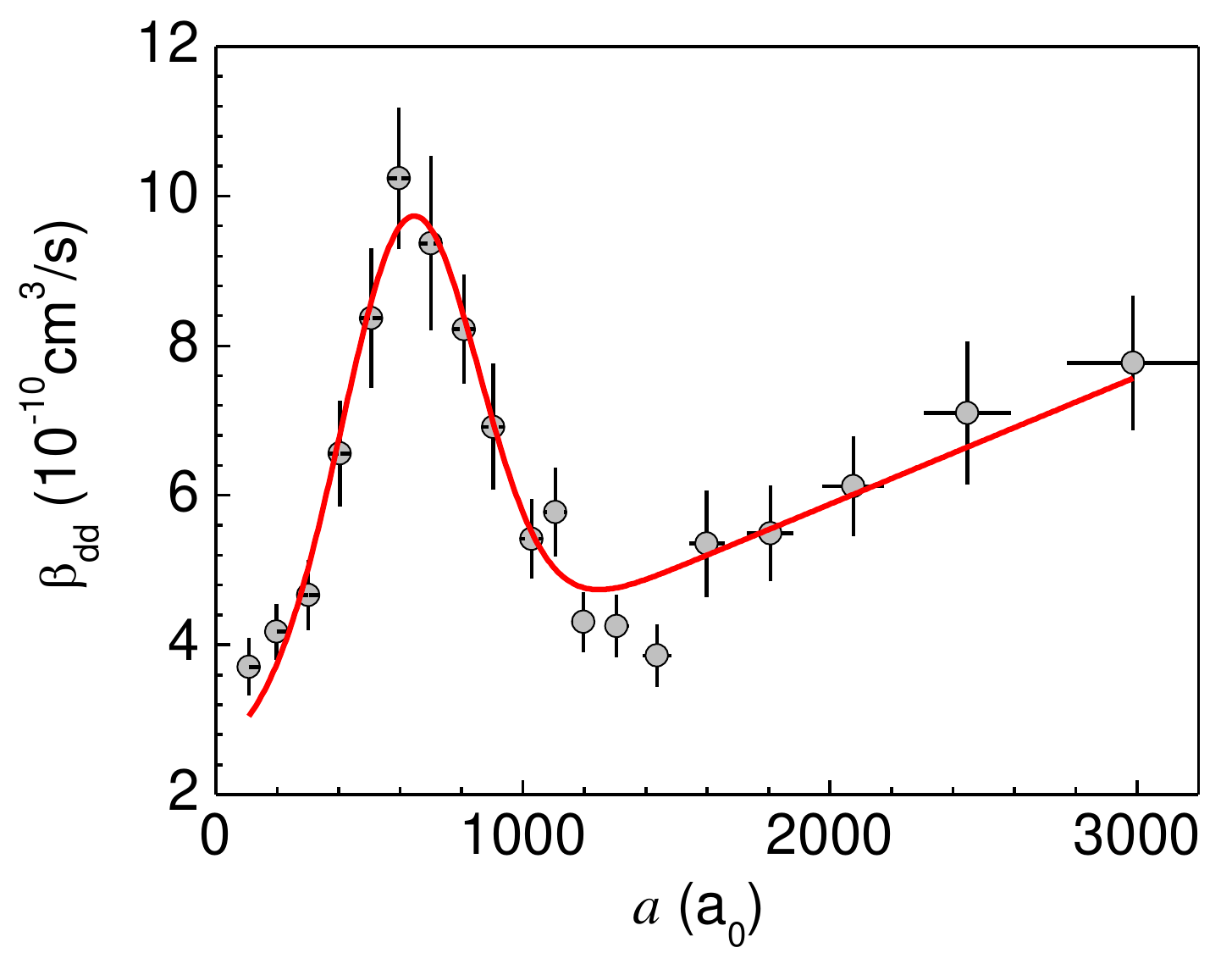}
\caption{\label{fig_dimerdimer} 
The filled circles show the dimer-dimer loss rate coefficient $\beta_{\rm{dd}}$ as a function of $a$ for NaRb Feshbach molecule collisions for a sample with $T=343(30)$~nK.
The solid line shows the fit result. All scattering lengths are calculated using the $B$-to-$a$ conversion given in Eq.~\hyperref[eq_sc]{\ref{eq_sc}}. }
\end{figure}

The experimentally determined $\beta_{\rm{dd}}$ shown in Fig.~\hyperref[fig_dimerdimer]{\ref{fig_dimerdimer}} displays a clear resonance when the Na-Rb scattering length $a$ is around $600a_0$. According to the energy diagram in Fig.~\hyperref[fig1]{\ref{fig1}}, this resonance can potentially be attributed to the intersection of two NaRb FMs and the weakly-bound Na$_2$Rb$_2$ tetramer at $a = a_{\rm{dd}}^*$ or to the intersection of two NaRb FMs and one of the heteronuclear trimers plus another atom, i.e., NaRb$_2$ + Na at $a = a_{\rm dd,Na}^{\rm reaction}$ or  Na$_2$Rb + Rb at 
$a = a_{\rm dd,Rb}^{\rm reaction}$. However, from the relative positions of these intersections and the fact that this dimer-dimer resonance is observed at relatively small Na-Rb scattering length, we believe that it is reasonable to attribute the loss feature near $600 a_0$ to the Na$_2$Rb$_2$ tetramer instead of either of the two rearrangement 
reactions NaRb+NaRb $\rightarrow$ NaRb$_2$+Na and NaRb+NaRb $\rightarrow$ Na$_2$Rb+Rb. 

We are not aware of calculations of the lineshape for loss of dimers consisting of heteronuclear bosons. Using an empirical fit to a Gaussian profile with linear background~\cite{Pires2014}, we find that the resonance position is located at $a_{\rm{dd}}^* = 650(15)a_0$. This is nearly $50\%$ larger than the value of $420a_0$ for the resonance tied to the weakly-bound tetramer state predicted by our finite-range model. The fact that both the experimental data and the theoretical model exhibit a dimer-dimer 
resonance, albeit at somewhat different scattering lengths, is very encouraging. The fact that the agreement is, at present, at the qualitative and not at the quantitative level is not surprising given that the experiment operates at finite temperature and that the finite-range model makes a number of simplifying assumptions regarding the four-body dynamics. 

For the scattering of homonuclear bosonic FMs, which features two dimer-dimer resonances tied to two weakly-bound tetramers, D'Incao~{\em{et al.}}~\cite{DIncao2009} calculated the loss lineshape. In this case, there exists one rearrangement reaction (two dimers go to a trimer and an atom), which lies very close to the resonance associated with the
extremely weakly-bound (second) tetramer state~\cite{DIncao2009}. In contrast, our theoretical model for the heteronuclear case supports only one Na$_2$Rb$_2$ tetramer state but there exist two rearrangement reactions. Despite of these differences, our observed dimer-dimer loss lineshape is, in the vicinity of $a_{\rm{dd}}^*$, very similar to that calculated for the resonance associated with the more strongly-bound (first) tetramer state for homonuclear bosonic FMs, which is also located at relatively small 
scattering length~\cite{DIncao2009}. As in the homonuclear case, $\beta_{\rm{dd}}$ decreases to a local minimum that is located at around $a=1200a_0$ and then increases again approximately linearly till the largest experimentally accessible scattering length of $a=3000a_0$ is reached. 

Our theoretical model suggests that the NaRb+NaRb $\rightarrow$ Na$_2$Rb+Rb rearrangement reaction occurs at scattering lengths $a$ that are experimentally accessible (Appendix~\hyperref[sec_theory]{\ref{sec_theory}}). However, the experimental data provide no clear evidence for the existence of a loss maximum that could be attributed to this rearrangement reaction. One explanation might be that the associated loss peak is, just as in the homonuclear case~\cite{DIncao2009}, not very pronounced due to finite temperature effects. Since our theoretical model is only qualitative, it is also possible that the rearrangement reaction occurs at a larger $a$ than predicted by our model. If this was the case, it would not currently be accessible in 
our experiment.     

\begin{figure}
\includegraphics[width=0.9\linewidth]{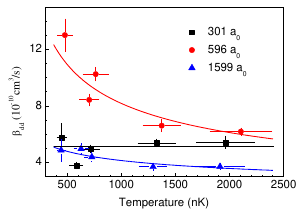}
\caption{\label{fig_varyT}
Temperature dependence of $\beta_{\rm{dd}}$ for $a=301 a_0$ (squares), $a=596 a_0$ (triangles) and $a=1599 a_0$ (circles). The black line is the mean of $\beta_{\rm{dd}}$ for the $301 a_0$ data. The red and blue lines show power law fits to the data for the other two scattering lengths. The horizontal error bars show the temperature uncertainty introduced by the heating during the hold time.}
\end{figure}

\section{Results: Temperature dependent behavior}
\label{sec_td}

Next we measure the temperature dependence of $\beta_{\rm dd}$ for different $a$, as shown in Fig.~\hyperref[fig_varyT]{\ref{fig_varyT}}. 
We focus on three cases: far away from the FR, near the resonance, and near the loss minimum. For $301 a_0$ (black squares), where the dimer is deeply bound with a binding energy of about 1~MHz, $\beta_{\rm dd}$ is nearly temperature independent with an average value of $5.1(8)\times 10^{-10}$~cm$^3$/s. This temperature independence is in accordance with the Wigner threshold law. However, the measured $\beta_{\rm dd}$ is larger than the calculated rate of $3.0(1)\times 10^{-10}$~cm$^3/s$ following the universal model~\cite{gao2010universal} with the $C_6$ coefficient of the NaRb-NaRb van der Waals interaction~\cite{Vexiau2015}. We have not yet investigated the reason for this discrepancy, but we note that a similar deviation was also observed for collisions between ground-state KRb molecules in the Wigner threshold regime~\cite{Ospelkaus2010}.

At $a = 596 a_0$, which is very close to the loss resonance, $\beta_{\rm dd}$ decreases with increasing temperature according to the power law $\beta_{\rm dd}\propto T^{-0.42(12)}$. At a much larger scattering length of $1599 a_0$, near the loss minimum, where the binding energy of the dimer is only 20~kHz, the extracted $\beta_{\rm dd}$ obeys a trend of $\beta_{\rm dd}\propto T^{-0.21(6)}$. These non-Wigner threshold law behaviors can be qualitatively explained by the near threshold effect studied in Ref.~\cite{Simbotin2014}. The Wigner threshold law holds when the collision energy is smaller than the smallest energy scale in the system. In the current case, this energy scale is set by the difference between the binding energy of two dimers and that of the tetramer. Near the dimer-dimer resonance, when this difference becomes smaller than the collision energy, the Wigner threshold law will be violated. Similar behavior was predicted to occur for the homonuclear case~\cite{DIncao2009, Deltuva2011}. 

\section{Conclusion}
\label{sec_con}
To conclude, we have studied inelastic collisions between ultracold heteronuclear NaRb FMs and observed a resonance tied to a Na$_2$Rb$_2$ tetramer state. This work is directly connected to the current interest in extending the Efimov scenario from homonuclear to heteronuclear, and from three- to higher-body cases, which eventually may lead to the quantitative understanding of a large class of few-body problems. While three-body Efimov resonances were already observed in several mixed-species systems~\cite{Barontini2009,Bloom2013,Hu2014,Pires2014,Tung2014,ulmanis2016heteronuclear,Maier2015,Wacker2016}, it has been realized that the few-body physics in mixtures is qualitatively different from that in homonuclear systems. Additional studies of bosonic and fermionic mixtures in the three- and higher-body sectors are thus needed to develop a solid understanding of how Efimov physics manifests itself in cold atom mixtures. The present work responded to this need and provided valuable information on 
the heteronuclear four-body sector, which had been studied poorly up to now.

Dimer-dimer resonances have many interesting applications. Just as atom-atom FRs allow one to control the effective atom-atom interaction strength, the dimer-dimer resonance
observed in this work can be used to control the effective interaction strength between two weakly-bound NaRb FMs. Scanning the $B$ field strength across the resonance at $a^*_{\rm{dd}}$, the dimer-dimer scattering length $a_{\rm{dd}}$ should change from positive to negative~\cite{DIncao2009,Deltuva2011}. Another interesting question is whether it is possible to create tetramers by magnetoassociation starting with a pure cloud of dimers in much the same way as dimers are being produced with atomic FRs. Although they have not yet been observed, the two rearrangement reactions may be useful in producing trappable Efimov trimers~\cite{DIncao2009}. 

\section{Acknowledgments}
We thank Bing Zhu, Bo Lu and Junyu He for laboratory assistance. This work was supported by the RGC General Research Fund (grants CUHK14301815 and CUHK14303317). DB gratefully acknowledges support by the National Science Foundation through grant number PHY-1806259. F.~W. and X.~Y. contributed equally to this work.

\appendix


\section{Theoretical framework}
\label{sec_theory}

\subsection{Overview of light-heavy Bose-Bose mixture}

It is useful to introduce the relevant system parameters. The two-body system parameters are the mass ratio $\kappa$, the light-light, heavy-heavy and light-heavy scattering lengths (denoted by $a_{\rm LL}$, $a_{\rm HH}$ and $a$), and the light-light, heavy-heavy and light-heavy van der Waals lengths. Either the scattering lengths at which the trimers hit the three-atom threshold or the binding energies at unitarity can be used to define the three-body parameters. For homonuclear systems, it was found that the three-body parameter is, with about 15\% accuracy, determined by the van der Waals length, i.e., the three-body parameter was found to be determined, to a good approximation, by two-body parameters~\cite{Wangjia2012,Naidon2014,Naidon2014b}. Reference~\cite{WangYuJun2012b,WangYuJun2012c} suggested that this notion could be extended to heteronuclear three-body van der Waals systems, in which the identical particles are either bosons or fermions; unfortunately, however, the emergent universality appeared to be rather complicated.

Recently, it was suggested that a zero-range theory based on the inter- and intraspecies two-body scattering lengths alone predicts a subset of the $^{133}$Cs$_2$-$^6$Li properties on the negative interspecies scattering length side qualitatively~\cite{Ulmanis2016}. Further, it was noticed that there exists a ``lower'' and an ``upper'' Efimov branch for this system~\cite{Ulmanis2016}. Application of this model to the positive interspecies scattering length side (see Appendix~\hyperref[appendix1]{\ref{appendix1}}) predicts that the atom-dimer scattering length $a_{\rm ad,Na}^*$ at which the lowest Na$_2$Rb Efimov trimer in the ``upper'' Efimov branch merges with the Na-NaRb atom-dimer threshold is larger than the atom-dimer scattering length $a_{\rm ad,Rb}^*$ at which the lowest NaRb$_2$ Efimov trimer in the upper Efimov branch merges with the Rb-NaRb atom-dimer threshold. The circles in Fig.~1 of the main text mark the scattering lengths at which enhanced losses due to the energetic degeneracy of the trimers and the NaRb dimer should be observable experimentally.

In the four-body sector, resonant loss features associated with processes that involve Na$_2$Rb$_2$, Na$_3$Rb, and NaRb$_3$ should, in principle, be observable. The latter two tetramers, Na$_3$Rb and NaRb$_3$, are expected to be tied to the Na$_2$Rb and NaRb$_2$ trimers, respectively. At unitarity, the existence of one LH$_3$ tetramer tied to each LH$_2$ trimer was predicted for a range of mass ratios, including the Rb-Na mass ratio of 3.78~\cite{Blume2019}.
This tetramer is expected to exist on the positive interspecies 
scattering length side till the tetramer energy 
hits the LH+H+H or LH$_2$+H thresholds. 
Moreover, near the dimer+atom+atom threshold, a new sequence of effective three-body Efimov states consisting of a tightly-bound light-heavy dimer and two atoms has been predicted to exist for mass ratios 30 and 50~\cite{WangYuJun2012}. Whether such effective three-body Efimov states exist for much smaller mass ratios is presently unknown. Last, whether Na$_3$Rb tetramers---attached to the so-called Efimov unfavored Na$_2$Rb trimer---exist, has, to the best of our knowledge, not yet been studied.

In this work, we focus on the Na$_2$Rb$_2$ tetramer, which can be probed experimentally via collisions between two NaRb dimers in an ultracold molecular NaRb sample. Theoretically, this four-body system is particularly interesting as it contains the NaRb$_2$ and Na$_2$Rb trimer subsystems. Thus, it is {\em{a priori}} not clear, if the Na$_2$Rb$_2$ tetramer is, at least predominantly, associated with one of the two sub-Efimov trimers. Our finite-range low-energy Hamiltonian, which is constructed to roughly reproduce the energies of the Na$_2$Rb and NaRb$_2$ trimers obtained within the zero-range framework (see Appendix~\hyperref[appendix2]{\ref{appendix2}} for details), predicts the existence of one Na$_2$Rb$_2$ tetramer that lies below the NaRb$_2$ trimer and, roughly, traces the energy of this trimer (see Fig.~1 
of the main text
for a schematic). The critical scattering length $a_{\rm dd}^*$ is found to be larger than the critical scattering length $a_{\rm ad,Rb}^*$. 
The next two subsections present the zero-range model 
for the three-body system (Appendix~\hyperref[appendix1]{\ref{appendix1}}) 
and the finite-range model, which treats the three- and four-body sectors (Appendix~\hyperref[appendix2]{\ref{appendix2}}).

\subsection{Three-body system with zero-range two-body interactions}
\label{appendix1}

\begin{figure}
	\includegraphics[width=0.9\linewidth]{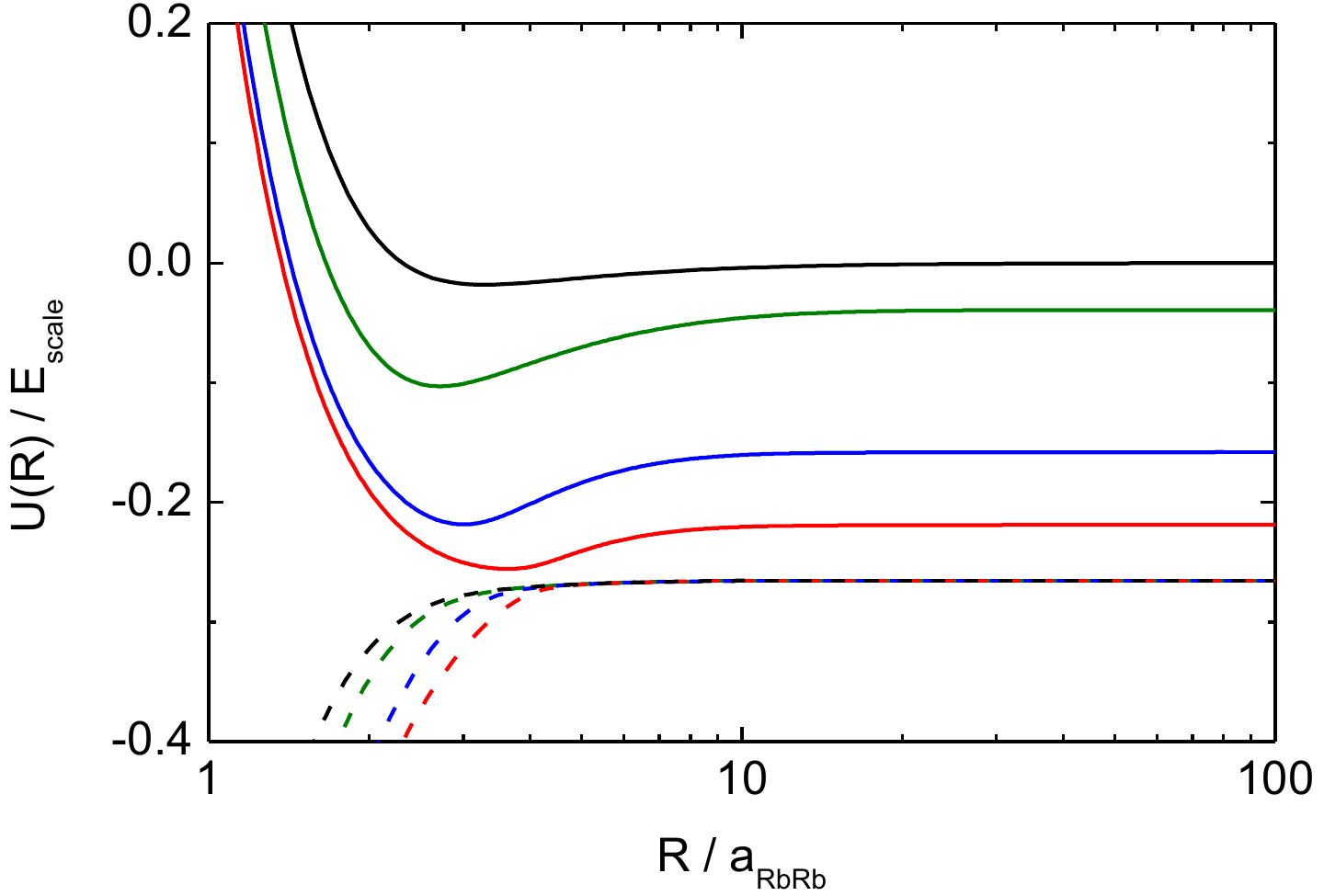}
	\caption{\label{fig1_appendix} 
		Effective adiabatic hyperspherical potential curves $U(R)$ for the NaRb$_2$ system with zero-range interactions. The scattering length ratios considered are $a /a_{\rm RbRb}=100$ (black), $4$ (green), $2$ (blue), and $1.7$ (red), from top to bottom for both the solid and the dashed lines. The potential curves shown by dashed and solid lines correspond to the lower Efimov branch and the upper Efimov branch, respectively. The dashed lines approach the Rb$_2$ dimer energy of $-\kappa^{-1}E_{\rm scale}=-0.2645E_{\rm scale}$ at large $R$. The solid lines approach the NaRb dimer energy of $-(1+\kappa^{-1}) (a_{\rm RbRb}/a)^2 E_{\rm scale}/2=-0.6323(a_{\rm RbRb}/a)^2 E_{\rm scale}$ at large $R$.}
\end{figure}

In our first set of calculations, the NaRb$_2$ system is treated in the adiabatic hyperspherical approximation~\cite{Macek1968,Klar1978,Starace1979} with two-body zero-range inter- and intraspecies interactions. Figure~\hyperref[fig1_appendix]{\ref{fig1_appendix}} shows the two lowest effective adiabatic potential curves as a function of the hyperradius $R$ for a fixed Rb-Rb scattering length and various interspecies scattering lengths $a$, namely $a/a_{\rm RbRb}=100,4,2$ and $1.7$. The hyperradius $R$ is defined as $R^2=d^{-2} (r_{13})^2 + d^{2} (r_{13,2})^2$, where $r_{13}$ denotes the distance between the Na atom (atom 3) and one of the Rb atoms (atom 1) and $r_{13,2}$ denotes the distance between the center of mass of the 13 subunit and the second Rb atom (atom 2). The mass scale $d$ is defined as $d^2=\mu_{13,2}/\mu_R$, where $\mu_{13,2}=(m_{\rm Rb}+m_{\rm Na}) m_{\rm Rb} /(2 m_{\rm Rb}+m_{\rm Na})$ is the reduced mass associated with the Jacobi distance $r_{13,2}$ and $\mu_R$ the hyperradial mass [$\mu_R^2 = m_{\rm Rb}^2 m_{\rm Na}/(2 m_{\rm Rb} + m_{\rm Na})$]. Fig.~\hyperref[fig1_appendix]{\ref{fig1_appendix}} scales the hyperradius by $a_{\rm RbRb}$ and the energy by $E_{\rm scale}=\hbar^2/[m_{\rm Na} (a_{\rm RbRb})^2]$. 

The lowest adiabatic potential curve (dashed lines) approaches the Rb$_2$ dimer energy at large $R$ for all $a$ and negative infinity as $R$ goes to zero, indicating that the Hamiltonian needs to be supplemented by a three-body parameter to avoid the Thomas collapse. The three-body states that ``live'' in these potential curves are referred to as belonging to the first or lower Efimov branch. The second lowest potential curve (solid lines), in contrast, approaches the NaRb dimer energy at large $R$ for all $a$. This NaRb dimer threshold lies above the Rb$_2$ threshold for the $a$ considered; it crosses the Rb$_2$ threshold at $a=1.546a_{\rm RbRb}$, i.e., at an interspecies scattering length that is somewhat smaller than those considered in Fig.~\hyperref[fig1_appendix]{\ref{fig1_appendix}}. The solid lines exhibit a minimum around $R \approx 3-4a_{\rm RbRb}$ and approach positive infinity as $R$ goes to zero. The three-body states that live in these potential curves are referred to as belonging to the second or upper Efimov branch. The repulsive small-$R$ behavior implies that one can, within the adiabatic hyperspherical approximation, calculate the NaRb$_2$ energies for the zero-range interaction model based solely on the two-body scattering lengths.

We calculate the three-body energies in the second lowest adiabatic potential curves and search for the scattering length at which the NaRb$_2$ trimer energy is equal to the NaRb dimer energy, i.e., we search for the scattering length ratio $a/a_{\rm RbRb}$ at which the adiabatic potential curves shown by solid lines in Fig.~\hyperref[fig1_appendix]{\ref{fig1_appendix}} cease to support a three-body bound state. We find that this occurs at $a = a_{\rm ad,Rb}^* \approx 1.72 a_{\rm RbRb}$ or, plugging in the Rb-Rb scattering length for the experimentally relevant resonance, at $a_{\rm ad,Rb}^* \approx 173a_0$. 

It should be kept in mind that the adiabatic hyperspherical approximation is, as the name indicates, an approximation. Inclusion of the adiabatic correction changes the critical scattering length prediction by about 5\%, i.e., we find $a_{\rm ad,Rb}^* \approx 1.81 a_{\rm RbRb}$ (we note that the percentage corrections are larger for the critical scattering lengths associated with excited states). While the relatively small change upon inclusion of the adiabatic correction may be interpreted as suggesting that the adiabatic hyperspherical approximation makes quantitative predictions, we cautiously note that these values cannot, since we are dealing with excited states, be interpreted as lower and upper bounds~\cite{Coelho1991}. In principle, the entire set of adiabatic potential curves and associated channel couplings should be taken into account. Such a calculation is, however, not pursued here. One of the reasons is that the zero-range approximation itself needs to be extended to account for finite-range effects. Our premise is that the adiabatic hyperspherical framework provides physical insights as well as estimates for the critical scattering lengths that can be used to qualitatively, and possibly semi-quantitatively, explain aspects of the experimental results.

In addition to the critical scattering length $a_{\rm ad,Rb}^*$ associated with the lowest NaRb$_2$ trimer state of the second Efimov branch, we used the adiabatic potential curves to calculate other critical scattering lengths. The next three higher-lying trimers hit the atom-dimer threshold at $a/a_{\rm RbRb} \approx 29$, $a/a_{\rm RbRb} \approx 1250$, and $a/a_{\rm RbRb} \approx 47000$, yielding for the scattering length ratios approximately $1/17$ for the lowest two states, $1/43$ for the second- and third-lowest states, and $1/38$ for the next pair of states. These values suggest that the finite value of the Rb-Rb scattering length has a profound effect on the scattering length ratios on the positive interspecies scattering lengths side. Qualitatively, this can be understood by realizing that the lowest trimer state at the atom-dimer threshold has three ``active'' interactions with approximately equal scattering length (the scaling factor for three resonant interactions is $\lambda=16.12$) while the higher-lying states at the atom-dimer threshold resemble more and more the situation where only two interactions are ``active'' (recall, the scaling factor for two resonant interactions is $\lambda_{\rm NaRbRb}= 37$).

For completeness, we also report the scattering lengths for the two lowest states for which the trimers hit the three-atom threshold on the negative interspecies scattering length side. The values are $a_{\text{Rb}}^- \approx -118 a_{\rm RbRb}$ and $ -4085 a_{\rm RbRb}$, yielding a ratio of $1/34.6$ and suggesting that the experimental observation of the first three-atom loss feature may already require rather good magnetic field control. In addition, there may exist three-atom loss features that are associated with three-body states that live in the lower Efimov branch; these are not considered here.

We also treated the Na$_2$Rb system in the adiabatic hyperspherical approximation. For the upper Efimov branch (this is the branch for which the corresponding adiabatic hyperspherical potential curves approach the NaRb dimer energy at large $R$), the lowest trimer hits the atom-dimer threshold at $a_{\rm ad,Na}^* \approx 2.8 a_{\rm RbRb}$ and the three-atom threshold at $a_{\rm Na}^- \approx -40300 a_{\rm RbRb}$. Combining the NaRb$_2$ and Na$_2$Rb results, the adiabatic hyperspherical framework predicts $a_{\rm ad,Na}^*/a_{\rm ad,Rb}^* \approx 2.8 / 1.72\approx 1.63$. We note that the critical scattering length $a_{\rm ad,Na}^*$ depends quite sensitively on corrections beyond the adiabatic hyperspherical approximation, suggesting that it is quite possible that the ratio $a_{\rm ad,Na}^*/a_{\rm ad,Rb}^*$ 
is larger than $1.63$.

The description of the three-body system could be made more quantitative by constructing a Hamiltonian that employs finite-range two-body model interactions with the correct scattering lengths and van der Waals tails. While this is a worthwhile avenue to pursue, solving the corresponding four-body Schr\"odinger equation is a rather challenging task that is not pursued here. Instead, the next section develops a simple finite-range framework for which the four-body Schr\"odinger equation can be solved fairly straightforwardly.

\subsection{Finite-range low-energy model}
\label{appendix2}

\begin{figure}
	\includegraphics[width=0.9\linewidth]{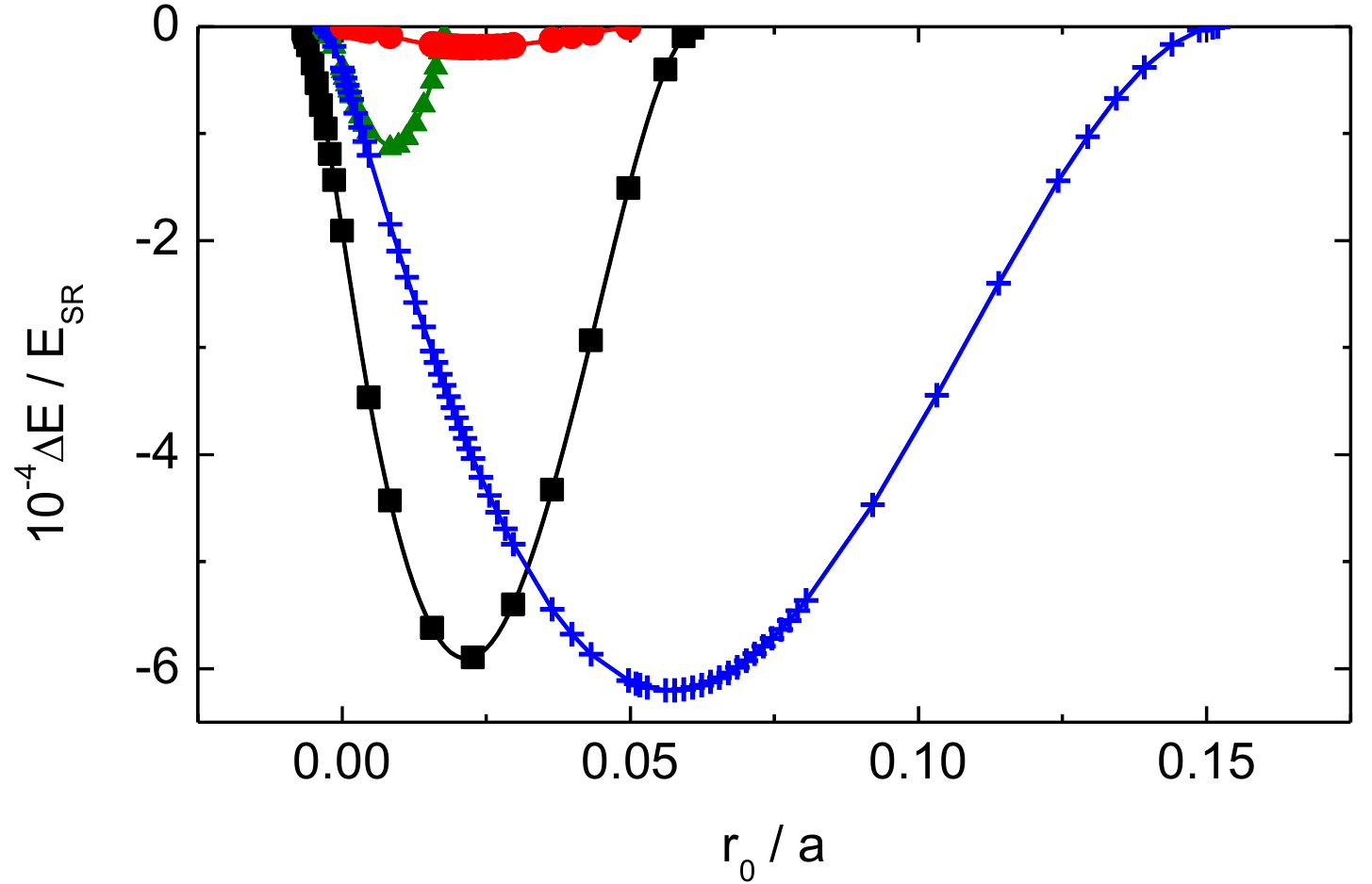} 
	\caption{\label{fig2_appendix}
		Three- and four-body energies, relative to various thresholds, calculated using the low-energy finite-range Hamiltonian as a function of $1/a$. The pluses show the energy difference $\Delta E=E_{{\rm{Na}}{\rm{Rb}}_2}-E_{\rm NaRb}$; on the positive $a$ side, the energy difference is zero at $a=a_{\rm ad,Rb}^*$. The circles show the energy difference $\Delta E=E_{{\rm{Na}}_2{\rm{Rb}}}-E_{\rm NaRb}$; on the positive $a$ side, the energy difference is zero at $a=a_{\rm ad,Na}^*$. The triangles show the energy difference $\Delta E=E_{{\rm{Na}}{\rm{Rb}}_2}-2 E_{\rm NaRb}$; on the positive $a$ side, the energy difference is zero at the critical scattering length where the rearrangement reaction NaRb+NaRb $\rightarrow$ NaRb$_2$+Na is enhanced. The squares show the energy difference $\Delta E=E_{{\rm{Na}}_2{\rm{Rb}}_2}-2E_{\rm NaRb}$; on the positive $a$ side, the energy difference is zero at $a=a_{\rm dd}^*$. The low-energy Hamiltonian predicts $a_{\rm dd}^* > a_{\rm ad,Rb}^*$. The energy differences are scaled by $E_{\rm SR}=\hbar^2/(2 \mu_{\rm NaRb}r_0^2)$, where $\mu_{\rm NaRb}$ is the reduced mass of the NaRb dimer.
	}
\end{figure}

To treat the Na$_2$Rb$_2$ and NaRb$_3$ tetramers, we employ a finite-range model that excludes both Rb$_2$ dimers and Na$_2$ dimers. As a consequence, the model describes the NaRb$_2$ and Na$_2$Rb trimers that live in the upper Efimov branch but not those that live in the lower Efimov branch.

The model assumes that each Rb-Na pair interacts through an attractive two-body Gaussian potential with fixed range $r_0$ and variable depth $v_0$; the depth $v_0$ is adjusted to dial in the desired value of the interspecies scattering length $a$. In addition, each Na-Rb-Rb triple interacts through a purely repulsive Gaussian three-body potential with range $R_0$ and height $V_{0,{\rm Rb}}$~\cite{Blume2014} (note this three-body interaction potential is distinct from the effective adiabatic potentials discussed in Appendix~\hyperref[appendix1]{\ref{appendix1}}). Similarly, each Na-Na-Rb triple interacts through a purely repulsive Gaussian three-body potential with the same range $R_0$ and height $V_{0,{\rm Na}}$. Throughout, the range $R_0$ is fixed. The height $V_{0,{\rm Rb}}$, which serves to set the energy scale of the NaRb$_2$ trimer, is also fixed and chosen such that the interspecies scattering length $a$ at which the NaRb$_2$ trimer energy hits the atom-dimer threshold is a few times larger than $r_0$ and $R_0$; this separation of scales ensures that the results are, to a good approximation, independent of the details of the model potentials. The height $V_{0,{\rm Na}}$, in turn, is varied. For each fixed $V_{0,{\rm Na}}$, the NaRb$_2$ trimer and Na$_2$Rb$_2$ tetramer energies are calculated as a function of $a$ and the critical scattering lengths $a_{\rm ad,Rb}^*$ and $a_{\rm dd}^*$ are determined. The strategy is then to choose the ``best'' $V_{0,{\rm Na}}$ such that the ratio  $a_{\rm ad,Rb}^*/a_{\rm ad,Na}^*$ is roughly the same as that for the zero-range model. The critical scattering length $a_{\rm dd}^*$ obtained in this manner is a prediction of this low-energy model Hamiltonian. Note that the critical scattering length at which the NaRb$_2$ trimer and the NaRb$_3$ tetramer become unbound is independent of $V_{0,\rm{Na}}$.

The idea behind the finite-range low-energy interaction model is that the repulsive Rb-Rb scattering length is accounted for, in an effective manner, by a purely repulsive three-body potential that introduces a repulsive short-range repulsion in the adiabatic potential curve for the NaRb$_2$ system. Similarly, the repulsive Na-Na scattering length is accounted for, again in an effective manner, by a purely repulsive three-body potential that introduces a repulsive short-range repulsion in the adiabatic potential curve for the Na$_2$Rb system. Adjusting $V_{0,{\rm Na}}$ while keeping $V_{0,{\rm Rb}}$ fixed then allows one to ``dial in'' the desired
relative strengths of these short-range repulsions. 
We refer to this model as a low-energy model since the  deeper-lying Na$_2$ and Rb$_2$ thresholds are not accounted for at all, excluding, e.g., the possibility that the Na$_2$Rb$_2$ tetramer breaks up into Na$_2$ and Rb$_2$.

We solve the time-independent Schr\"odinger equation for the low-energy Hamiltonian by a basis set expansion approach, namely, we use explicitly correlated Gaussian basis functions with non-linear parameters that are optimized semi-stochastically~\cite{Suzuki1998,Mitroy2013}. Figure~\hyperref[fig2_appendix]{\ref{fig2_appendix}} shows the three- and four-body energies---with the threshold energies subtracted---for our finite-range model. Pluses show the energy difference $E_{\rm{NaRb}_2}-E_{\rm NaRb}$ as a function of $r_0/a$ while triangles show the energy difference $E_{\rm{NaRb}_2}-2E_{\rm NaRb}$ as a function of $r_0/a$, where $E_{\rm{NaRb}_2}$ is the lowest NaRb$_2$ trimer energy of our model Hamiltonian. The points at which these energy differences vanish are the critical scattering length values. On the negative scattering length side, this is the critical scattering length $a_{\rm Rb}^-$ and on the positive scattering length side, these are the critical scattering lengths $a_{\rm ad,Rb}^*$ 
(solid circle in Fig.~1 of the main text)
and $a_{\rm dd,Na}^{\rm reaction}$, corresponding to the square in 
Fig.~1 of the main text (i.e., the scattering length at which the rearrangement reaction $\rm{NaRb}+\rm{NaRb} \rightarrow \rm{NaRb}_2 +\rm{Na}$ is expected to be enhanced). The model predicts the scattering length ratio $a_{\rm dd,Na}^{\rm reaction}/a_{\rm ad,Rb}^* \approx 8$ (in the zero-range model, this ratio is about 10 in the hyperspherical approximation and 7 if the adiabatic correction is included) and a ratio of about 20 for the atom-dimer resonance positions of the two lowest states (recall, the zero-range model, treated in the adiabatic approximation, yielded $\approx 17$). These comparisons show that our finite-range model reproduces the NaRb$_2$ trimer properties on the positive scattering length side predicted by the zero-range model quite well.

Our potential model predicts,
in agreement with Ref.~\cite{Blume2019},
the existence of exactly one NaRb$_3$ tetramer state. The energy of this state is not shown in Fig.~\hyperref[fig2_appendix]{\ref{fig2_appendix}}. Within our numerical accuracy, the NaRb$_3$ tetramer becomes unbound at the same interspecies scattering length as the NaRb$_2$ trimer.

\begin{table}
	\caption{
		Summary of the calculated few-body resonances in the Na-Rb system. The results in the column labeled ``zero range'' are obtained using the zero-range two-body interaction model introduced in Appendix~\hyperref[appendix1]{\ref{appendix1}} while those in the column labeled ``finite range'' are obtained using the finite-range low-energy model introduced in Appendix~\hyperref[appendix2]{\ref{appendix2}}.}
	\begin{tabular}{c|c|c}
		\hline\hline
		resonances & zero range & finite range\\
		\hline\hline
		$a^*_{\rm ad,Rb}$&	$173a_0$ &  \\
		\hline
		$a^*_{\rm ad,Rb,1}$&	$2941a_0$ & $3460a_0$  \\
		\hline
		$a^*_{\rm ad,Na}$&  $281a_0$ &  \\  		 
		\hline
		$a^*_{\rm dd}$ & & $\sim 420a_0$  \\
		\hline
		$a^{\rm reaction}_{\rm dd,Na}$ & $1730a_0$ & $1384a_0$   \\
		\hline	
		$a^-_{\rm Rb}$ & & $\sim -11850a_0$  \\ 
		\hline
		$a^-_{\rm Na}$ & & $\sim -4.05\times 10^6a_0$  \\ 	
		\hline
	\end{tabular}
	\label{table1}
\end{table}

Next we consider the Na$_2$Rb trimer and Na$_2$Rb$_2$ tetramer, whose energies depend, within our model, on the three-body height $V_{0,\rm{Na}}$. The circles in Fig.~\hyperref[fig2_appendix]{\ref{fig2_appendix}} show the energy difference $E_{\rm{Na}_2\rm{Rb}}-E_{\rm NaRb}$ for the height $V_{0,\rm{Na}}$ of the repulsive Na-Na-Rb potential chosen such that $a_{\rm ad,\rm{Na}}^*$ is about three times larger than $a_{\rm ad,Rb}^*$. For this model Hamiltonian, the energy difference $E_{\rm{Na}_2\rm{Rb}_2}-2 E_{\rm NaRb}$ (squares in Fig.~\hyperref[fig2_appendix]{\ref{fig2_appendix}}) goes to zero at $a=a_{\rm dd}^* \approx 2.45 a_{\rm ad,Rb}^*$. If  $V_{0,\rm{Na}}$ is chosen such that $a_{\rm ad,\rm{Na}}^*$ is about two times larger than $a_{\rm ad,Rb}^*$, we find $a=a_{\rm dd}^* \approx 2.17 a_{\rm ad,Rb}^*$ (recall, the zero-range model in the adiabatic approximation predicts $a_{\rm ad,Na}^*/a_{\rm ad,Rb}^*=1.63$; the actual value, however, is---as discussed in Appendix~\hyperref[appendix1]{\ref{appendix1}}---expected to be somewhat larger). This shows that $a_{\rm dd}^*$ depends, if expressed in terms of $a_{\rm ad,Rb}^*$, less strongly on the value of $V_{0,\rm{Na}}$ than $a_{\rm ad,Na}^*$, suggesting that the Na$_2$Rb$_2$ tetramer properties are primarily determined by the properties of the NaRb$_2$ trimer. Our calculations suggest that $a_{\rm dd}^*$ is larger than $a_{\rm ad,Rb}^*$. We cannot determine unambiguously whether  $a_{\rm dd}^*$ is greater or smaller than $a_{\rm ad,Na}^*$, primarily because the zero-range prediction for $a_{\rm ad,Na}^*$ depends sensitively on whether or not the adiabatic correction is included. Our potential model, using what we consider reasonable values for $V_{0,\rm{Na}}$, predicts the existence of exactly one Na$_2$Rb$_2$ tetramer state. 
The calculated resonance positions for the zero-range and finite-range 
models are summarized in Table~\hyperref[table1]{\ref{table1}}.
\bibliographystyle{apsrev4-1}

\end{document}